\documentstyle{mn}

\part*{}

\title[Turbulence in protostellar envelopes]{Turbulence in Class 0 
and Class I protostellar envelopes}

\author[Ward-Thompson, Hartmann \& Nutter]{D. Ward-Thompson$^1$, 
L. Hartmann$^2$, D. J. Nutter$^1$ \\
$^1$Department of Physics and Astronomy, Cardiff University, 
PO Box 913, Cardiff, CF24 3YB \\
$^2$Harvard-Smithsonian Center for Astrophysics, 60 Garden Street, 
Cambridge, MA02138, USA \\}
\date{Accepted 2004 November 18; received 2004 November 17; in original 
form 2003 October 17.} 

\pagerange{000--000} 

\begin{document}

\label{firstpage}

\maketitle

\begin{abstract}
We estimate the levels of turbulence in the envelopes of Class 0
and I protostars using a model based on measurements of
the peak separation of double-peaked asymmetric line profiles.
We use observations of 20 protostars of both Class 0 \& I taken in the 
HCO$^+$(J=3$\rightarrow$2) line that show the classic double-peaked
profile. We find that some Class 0 sources show high levels of turbulence
whilst others demonstrate much lower levels. In Class I protostars
we find predominantly low levels of turbulence. The observations are 
consistent with a scenario in which Class 0 protostars form in a variety of
environments and subsequently evolve into Class I protostars. 
The data do not
appear to be consistent with a recently proposed scenario in which 
Class 0 protostars can only form in extreme environments.
\end{abstract}

\section{Introduction}

A model for the evolution of the youngest protostars has emerged over the
last decade or more, based on a great deal of observational evidence. In
this model the protostar evolves through a series of recognisable stages
in which a circumstellar envelope of material collapses onto a central
protostar and its surrounding disc (see Andr\'e, Ward-Thompson \&
Barsony 2000 for a review).

In this scenario the first recognisable stage of protostellar evolution 
is known as the Class 0 stage (Andr\'e, Ward-Thompson \& Barsony 1993 -- 
hereafter AWB93), which is defined in such a way that the circumstellar
envelope contains more mass than the central protostar. The next stage is
known as the Class I stage (Lada \& Wilking 1984; Lada 1987), 
in which there is still a massive envelope, although there is more mass
in the central protostar. The subsequent Class II \& III stages (Lada \& 
Wilking 1984; Lada 1987) represent protostars whose circumstellar
envelopes have been dissipated, but in which there is still a
circumstellar disc (e.g. Kenyon \& 
Hartmann 1995; Safier, McKee \& Stahler 1997). The Class II \& III stages
are also known as the classical T Tauri (CTT) and weak-line T Tauri (WTT)
stages respectively (Andr\'e \& Montmerle 1994).

Infall from the protostellar envelope onto the protostar can be inferred
from the double-peaked asymmetric spectral line profiles
of certain molecular species (e.g. Zhou 1992; 1995; Rawlings et al. 1992;
Zhou et al. 1993; Ward-Thompson et al. 1996). 
There are a number of explanations for the manner in
which the asymmetry arises, but the usually accepted explanation is based
on the observer preferentially seeing
warmer blue-shifted material closer to the protostar and 
cooler red-shifted material further from the protostar 
(Zhou 1992). Infall profiles have now been observed towards many
protostars (e.g. Gregersen et al. 1997; 2000).

Our understanding of
the evolutionary process has been formalised into a luminosity
versus colour diagram analogous to the HR diagram, using the bolometric
luminosity of the protostar, L$_{{\rm Bol}}$, and a `bolometric temperature',
T$_{{\rm Bol}}$ (Myers \& Ladd 1993). The bolometric temperature is based
on the position of the peak wavelength of the continuum emission from
the protostar. The more evolved the protostar, the greater its measured
value of T$_{{\rm Bol}}$. Hence T$_{{\rm Bol}}$ provides a useful 
evolutionary `age' indicator for protostars.

\begin{figure*}
\setlength{\unitlength}{1mm}
\begin{picture}(80,40)
\includegraphics{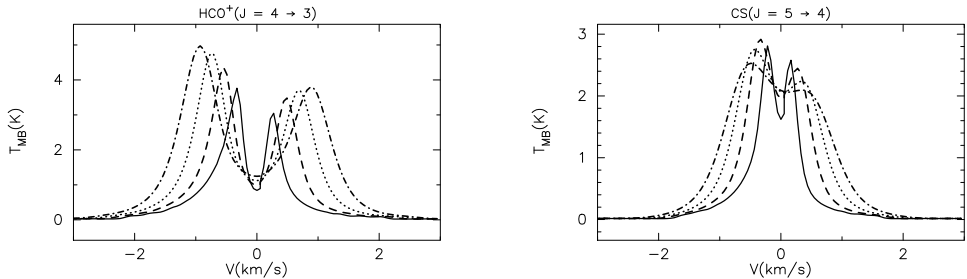}
\end{picture}
\caption{Plots showing the dependence of the predicted 
HCO$^+$(J=4$\rightarrow$3) and CS(J=5$\rightarrow$4) line profiles on the
magnitude of the turbulent velocity dispersion in the protostellar
envelope (assumed to be constant throughout the 
envelope). Spectra are plotted for turbulent velocity dispersions of 0.1~km/s
(solid lines), 0.2~km/s (dashed lines), 0.3~km/s (dotted lines) and 0.4~km/s 
(dot-dashed lines). The equivalent FWHM turbulent velocity widths are 
found by multiplying these numbers by 2.35.
Note that increasing turbulence increases the peak separation.
The same effect is also seen in other transitions, such as
HCO$^+$(J=3$\rightarrow$2).}
\end{figure*}

Recent work based on chemical modelling of protostellar envelopes
appears to support the idea that Class I sources are more evolved
than Class 0 sources (Buckle \& Fuller 2003). These authors modelled
the chemical evolution of sulphur-bearing molecules and showed that
Class I protostellar envelopes are more chemically evolved in general
than the envelopes of Class 0 protostars.

However, the role of turbulence in the star formation process is
still a matter for debate (e.g. Elmegreen 2002).
In this paper we endeavour to measure the relative amounts 
of turbulence in Class 0 \& I protostellar envelopes
by measuring the
peak separation in a sample of protostars. Furthermore, we interpret
the variation in separations observed in terms of our recent radiative
transfer model results. In this model the velocity separation of the
peaks is seen to be dominated by the relative level of
turbulence in the infalling protostellar envelope.

\section{Measuring turbulence}

The presence of non-thermal motions in molecular cloud regions where
stars are formed has been recognised for some time (e.g. Caselli \& 
Myers 1995). These non-thermal motions are usually attributed to the
presence of turbulence (e.g. Padoan \& Nordlund 2002). Measuring 
the exact levels of turbulence in molecular clouds can be complicated.
However, the fact that turbulence plays a significant role in the
star formation process is now widely recognised (e.g. Elmegreen 2002
and references therein).

We have recently carried out detailed spectral modelling of
protostellar infall candidates (Ward-Thompson \& Buckley 2001). This process
uses a radiative transfer $\Lambda$-iteration model based on the method
of Stenholm (1977). The method was refined by subsequent workers
(Matthews 1986; Heaton et al. 1993; Buckley 1997). It
solves the spectral line radiative
transfer problem for the rotational transitions of linear molecules in
a spherically symmetric model cloud. The radial profiles
of infall velocity, temperature, density, tracer molecule 
abundance and micro-turbulent velocity dispersion may be specified
(Ward-Thompson \& Buckley 2001).

The method is begun by choosing an initial 
radiation field in a more or less arbitrary manner. From this,
a `false' set of molecular energy level populations
are calculated. Radiative transitions between these populations will 
generally produce a radiation field which departs from the initial field.
If this radiation field is used to calculate a new set of
populations, and the procedure is repeated a sufficiently large number
of times, the radiation field and populations should
eventually converge on a mutually consistent
solution (for further details see Ward-Thompson \& Buckley 2001).
The output spectra are convolved with a chosen beam size to match
any given set of observations.
We have applied the method to HCO$^+$ and CS spectral line observations of
protostellar envelopes and modelled their infall parameters. 

Figure 1 shows the predicted line profiles that would be observed
at the James Clerk Maxwell Telescope (JCMT) in the HCO$^+$(J=4$\rightarrow$3)
and CS(J=5$\rightarrow$4) molecular line transitions to match some of the
data we show below. The beam convolution to match the
Caltech Submillimeter Observatory (CSO) data also shown below does
not significantly alter the results.
The model assumes an inside-out collapse in which an expansion
wave has reached a radius that we call the infall radius,
outside of which the velocity is zero.

We assume that a mass M$_{\odot}$/2 has already accreted onto the 
central protostar, and a further 
M$_{\odot}$/2 of envelope gas is infalling towards it. 
We choose an effective sound speed 
of $a_{\rm eff}=0.35\,{\rm km\, s}^{-1}$.
We use model relations for the 
radial velocity and density profiles
consistent with the inside-out collapse scenario of $\rho\propto r^{-3/2}$
inside the infall radius and $\rho\propto r^{-2}$
outside the infall radius.
We truncate the density profile at an outer radius of 10000\,AU, which 
encloses a total mass of 2.75\,M$_{\odot}$.
The temperature profile 
in the optically thin part of the envelope is 
chosen to have a canonical profile $T\propto r^{-0.4}$.
The parameter normalisations used in the radiative transfer 
modelling are given by Ward-Thompson \& Buckley (2001) and
the molecular constants are taken from the catalogue
of Poynter \& Pickett (1985). 

\begin{figure*}
\setlength{\unitlength}{1mm}
\begin{picture}(120,40)
\includegraphics{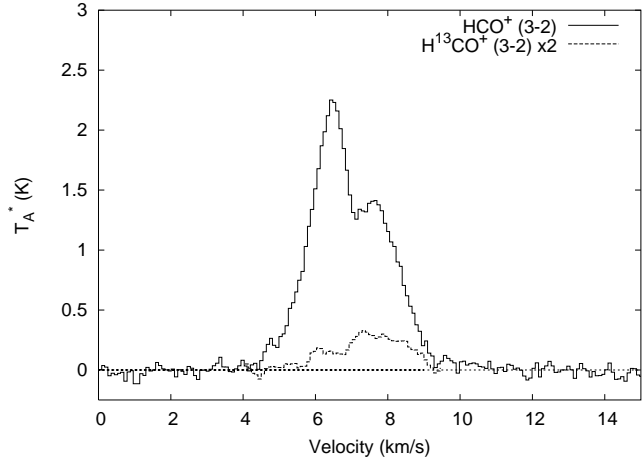}
\end{picture}
\vspace*{2cm}
\caption{HCO$^+$(J=3$\rightarrow$2) spectrum of L1489 showing the
double-peaked asymmetric line profile (solid line). Also shown is
the H$^{13}$CO$^+$(J=3$\rightarrow$2) isotope line (dashed line),
multiplied by 2, illustrating that the optically thin isotope is
single-peaked.}
\end{figure*}

\begin{table}
\caption{Summary of all of the data used in our analysis, listed in
order of increasing bolometric temperature.
Column 1 lists the usual source name, column 2 lists the bolometric
temperature, T$_{{\rm Bol}}$, column 3 indicates whether it is a
Class 0 or I protostar, column 4 gives the separation between
the two peaks of the asymmetric line profile (see text for details).
Column 5 contains the following notes:  (1) Our data; (2) data taken from 
Gregersen et al. (1997); (3) data taken from Gregersen et al. (2000).}
\begin{center}
\begin{tabular}{lcccc}
\hline
Name & T$_{{\rm Bol}}$ & Class & Separation & Notes \\ 
 & (K) & & (km/s) & \\
\hline
B335 & 29 & 0 & 0.78 & 2\\
NGC1333 IRAS4A & 34 & 0 & 1.70 & 2\\
Serpens SMM4 & 35 & 0 & 2.17 & 2\\
NGC1333 IRAS4B & 36 & 0 & 1.53 & 2\\
Serpens SMM2 & 38 & 0 & 1.23 & 2\\
L1527 & 41 & 0 & 0.85 & 2\\
L1157 & 44 & 0 & 0.97 & 2\\
Serpens SMM1 & 45 & 0 & 2.48 & 2\\
B228 & 48 & 0 & 0.55 & 3\\
NGC1333 IRAS2 & 52 & 0 & 1.57 & 3\\
L1448N & 55 & 0 & 0.90 & 1\\
CB244 & 56 & 0 & 0.87 & 3\\
IRAS 23011 & 57 & 0 & 1.01 & 3\\
IRAS 20050 & 69 & 0 & 2.03 & 2\\
L1634 & 77 & I & 0.71 & 3\\
IRAS 04166 & 91 & I & 0.72 & 3\\
L1251B & 91 & I & 1.36 & 3\\
IRAS 03235 & 136 & I & 0.73 & 3\\
TMC 1A & 170 & I & 0.61 & 3\\
L1489 & 238 & I & 1.10 & 1\\
\hline
\end{tabular}
\end{center}
\end{table}

Figure~1 shows how the line profiles 
depend on the magnitude of the turbulent velocity
dispersion, $\sigma_{\rm tb}$, when it is assumed 
to be uniform throughout the envelope. 
As the turbulent velocity dispersion increases, 
the most apparent effect on the line profile
is to increase the velocity separation between the 
two peaks. Figure~1 shows the results for HCO$^+$(J=4$\rightarrow$3), 
and CS(J=5$\rightarrow$4). A similar
result is also seen in the HCO$^+$(J=3$\rightarrow$2) transition.

The magnitude of the change 
in the line-shape observed in Figure~1 is totally different to that
caused by increasing the optical depth. Myers et al. (1996) explored 
the manner
in which increasing the optical depth changed the line-shape (see their 
figure~2a). They plotted the two extremes of optical depth that can cause a
double-peaked profile to be observed, and found that they
produce an increase of peak separation of only 
$\leq$0.1~kms$^{-1}$. Likewise, we find from our model that 
an increase of $\sim$2 in optical depth results in an
increase of peak separation of only 
$\sim$0.2~kms$^{-1}$ (c.f. figures 12 and 13 of Ward-Thompson
\& Buckley 2001). These two numbers can be
compared to the differences of order $\sim$1--2~kms$^{-1}$
caused by increasing turbulence, as seen in Figure 1.
Thus it can be seen that variations in optical depth will
only introduce a small scatter in the observed peak separation, which
is dominated by variations in levels of turbulence.

Hence we see that the relative level of turbulence in protostellar
envelopes can be estimated from the relative separation of the two
peaks of the classic double-peaked asymmetric infall profile. 
No other parameter has such a large effect on the
separation of the two peaks (Ward-Thompson \& Buckley 2001). 
Even though the profiles in Figure 1 are calculated for the
specific parameters listed above, the same effect of increasing
peak separation with increasing turbulence was seen in all models.
We now use this diagnostic to estimate the relative
levels of turbulence in the envelopes of protostars.

\section{Data}

The majority of the data we use are taken from the literature, but
data for two additional sources were obtained. L1489 was observed at
a position of R.A. (1950) = 04$^h$ 01$^m$ 40.6$^s$, Dec. (1950) =
$+$26$^\circ$ 10$^\prime$ 49$^{\prime\prime}$ and L1448N was observed at
a position of R.A. (1950) = 03$^h$ 22$^m$ 31.8$^s$, Dec. (1950) =
$+$30$^\circ$ 34$^\prime$ 45$^{\prime\prime}$.
These observations were carried out at the James Clerk Maxwell Telescope
(JCMT) on 2001 August 10, using the receiver
RxA2 (Davies et al. 1992), with the Digital Autocorrelation
Spectrometer (DAS) backend.
The DAS was configured with the optimum frequency 
resolution of 95kHz per channel, equivalent to $\sim 0.1$\,km\,s$^{-1}$.
This is very similar to the resolution of the literature data that we use.
The observations were made using frequency switching mode (Matthews 1996).
All data reduction was carried out using the {\small SPECX} package 
(Padman 1990). 

\begin{figure*}
\setlength{\unitlength}{1mm}
\begin{picture}(120,45)
\includegraphics{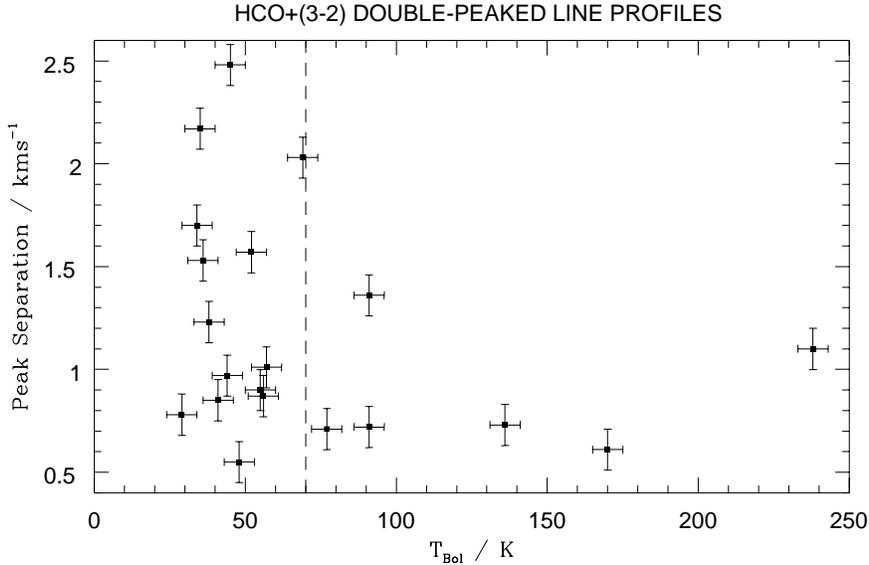}
\end{picture}
\vspace*{3cm}
\caption{Graph of peak separation versus T$_{{\rm Bol}}$ for Class 0 \& I
protostars. Large peak separations are equivalent to high levels
of turbulence. Higher values of T$_{{\rm Bol}}$ correspond to more evolved
protostars in the evolutionary scenario. The vertical dashed line
represents the border between Class 0 and Class I. Note that there
are both high and low levels of turbulence observed in Class 0
protostars.}
\end{figure*}

Figure 2 shows a typical result for one of the HCO$^+$(J=3$\rightarrow$2)
spectra. We see the two peaks of the line profile, separated by an
absorption dip, and the characteristic infall asymmetric profile.
From such a spectrum the positions of the red and blue peaks can be
obtained to an accuracy of roughly one spectral channel, or
$\sim 0.1$\,km\,s$^{-1}$.

We have spectral line data for a total of 20 sources --
14 Class 0 protostars and 6 Class I protostars -- in transitions of
HCO$^+$(J=3$\rightarrow$2) and H$^{13}$CO+(J=3$\rightarrow$2).
It is necessary to obtain data in the optically thin isotope line
to check that the two peaks being observed in the main isotope line
are not due to two different clouds along the same line of sight.
We could confirm that the rare isotope line was 
single-peaked in each case. The data are listed in Table 1.
All 20 sources have been well studied previously, and have known
continuum spectra and bolometric temperatures.
Table 1 lists the source name,
its bolometric temperature and Class. Table 1 also lists the measured
peak separation of the double-peaked spectra and the source of the
literature data.

\section{Discussion and Conclusions}

Figure 3 shows a plot of peak separation versus T$_{{\rm Bol}}$ 
for the sources
listed in Table 1. As shown above, peak separation scales with turbulence
level. Consequently those sources showing high levels of turbulence appear
in the upper parts of the graph, while those with lower levels of
turbulence appear in the lower parts of the graph.
The error-bars correspond to $\pm$5K and $\pm$0.1kms$^{-1}$.
T$_{{\rm Bol}}$ is used 
simply as a way of separating the Class 0 protostars from
the Class I protostars. The definition of a Class 0 protostar corresponds to 
it having T$_{{\rm Bol}}<$70K. The vertical dashed line is the dividing line 
between Class 0 and I protostars.

In Figure 3 we see Class 0 protostars with
a wide variety of levels of turbulence in their envelopes,
although we see that there are predominantly low levels
of turbulence in Class I sources. 
Figure 3 is consistent with the
evolutionary scenario proposed by AWB93, in which
Class 0 sources can form in regions of both high and low turbulence,
before evolving into Class I protostars.

One interesting facet of the data is that
there do not appear to be any Class I protostars in Figure 3
with high turbulence levels.
This may not be statistically significant, due to the smaller number
of Class I sources. However, if future data prove that this is
also significant, then this appears to be saying that any
initial turbulence in the environment of a forming protostar
dissipates before it
has evolved to the Class I stage. The more massive envelopes
in the earlier Class 0 protostars can sustain relatively high initial
levels of turbulence when they occur. However, by the Class I stage
when more than half of the envelope has accreted onto the central
protostar, a large percentage of any initial turbulence has been 
dissipated. 

We independently
estimated the line optical depths for the data in Table 1
(using the isotope line data), and
confirmed that they lie in the range $\sim$1.5--3.5, consistent with
the regime of parameters that we are modelling (see discussion in section
2 above). We plotted optical depth versus peak separation
and found no correlation. This appears to
confirm our assertion that the peak separation is not dominated by
optical depth effects, but rather by turbulence.

We note finally that an alternative evolutionary scenario for the
Class 0 \& I protostellar stages has recently been proposed
(Jayawardhana, Hartmann \& Calvet, 2001 -- hereafter JHC)
in which Class 0 protostars only form in high density environments
whilst Class I protostars form in low density environments, and the
two stages are in fact parallel evolutionary tracks. In this scenario
the high density environments required to produce Class 0 protostars
are produced by invoking colliding turbulent flows (see discussion in 
section 4.3 of JHC). Under such a scenario one might expect that this
additional turbulence required to produce Class 0 protostars would
be measurable, and that all Class 0 sources should exhibit high levels
of turbulence. Figure 3 shows this not to be the case, and therefore 
appears to be inconsistent with the JHC scenario.

\section*{Acknowledgments}

The authors would like to thank the staff of the JCMT
for assistance while these data were obtained.
JCMT is operated by the Joint Astronomy Centre, Hawaii, on
behalf of the UK PPARC, the Netherlands NWO, and the Canadian NRC.
DJN acknowledges PPARC for PDRA support.

\end{document}